\documentclass[reprint,footinbib,superscriptaddress,floatfix, aps, longbibliography]{revtex4-2}

\usepackage{graphicx}
\usepackage{dcolumn}
\usepackage{bm}
\usepackage{color}
\usepackage{soul}
\usepackage{xpatch}
\usepackage{empheq}
\usepackage{amsmath}
\usepackage{amssymb}
\usepackage{siunitx}
\usepackage[usenames,dvipsnames,svgnames,table]{xcolor}
\usepackage[colorlinks,linkcolor=blue,citecolor=blue,urlcolor=blue]{hyperref}
\usepackage[normalem]{ulem}
\usepackage[english]{babel}

\begin{document}

\newcommand{\jdp}[1]{\textcolor{teal}{#1}}

\title{Slow and fast topological dynamical phase transitions in a Duffing resonator driven by two detuned tones}

\author{Letizia Catalini}
\affiliation{Laboratory for Solid State Physics, ETH Z\"{u}rich, 8093 Z\"urich, Switzerland}
\affiliation{Quantum Center, ETH Z\"{u}rich, 8093 Z\"{u}rich, Switzerland}
\affiliation{Center for Nanophotonics, AMOLF, 1098XG Amsterdam, The Netherlands}
\email{L.Catalini@amolf.nl}

\author{Javier del Pino}
\affiliation{Department of Physics, University of Konstanz, 78464 Konstanz, Germany}
\author{Soumya S Kumar}
\affiliation{Department of Physics, University of Konstanz, 78464 Konstanz, Germany}
\author{Vincent Dumont}
\affiliation{Laboratory for Solid State Physics, ETH Z\"{u}rich, 8093 Z\"urich, Switzerland}
\affiliation{Quantum Center, ETH Z\"{u}rich, 8093 Z\"{u}rich, Switzerland}
\author{Gabriel Margiani}
\affiliation{Laboratory for Solid State Physics, ETH Z\"{u}rich, 8093 Z\"urich, Switzerland}
\affiliation{Quantum Center, ETH Z\"{u}rich, 8093 Z\"{u}rich, Switzerland}
\author{Oded Zilberberg}
\affiliation{Department of Physics, University of Konstanz, 78464 Konstanz, Germany}
\author{Alexander Eichler}
\affiliation{Laboratory for Solid State Physics, ETH Z\"{u}rich, 8093 Z\"urich, Switzerland}
\affiliation{Quantum Center, ETH Z\"{u}rich, 8093 Z\"{u}rich, Switzerland}

\begin{abstract}
Nonlinear dynamics are studied in diverse fields as climate models, avalanches, nanomechanical sensors, optical frequency converters, and electrical quantum amplifiers. A widely studied nonlinear model is the so-called Duffing (or Kerr) resonator, which features a quartic potential term. Two hallmark properties of this model are (i)~a shift of a system's resonance frequency as a function of the driving strength, and (ii)~monostable or bistable responses, depending on the drive strength and detuning from resonance. Together, these two properties can lead to dynamical phase transitions when several drives are applied simultaneously. Here, we report an experimental and theoretical study of a driven-dissipative nonlinear system with two detuned drives. We observe distinct response regimes characterized by the system's ability to follow the system's time-dependent vector-flow topology. Our work provides an example for understanding dynamical phase transitions in out-of-equilibrium nonlinear systems.
\end{abstract}

\maketitle

\section{Introduction}

Measuring the response of a system to a strong `pump' by means of a weak `probe' has been a cornerstone of optical spectroscopy and microscopy for many years~\cite{owyoung1978coherent,lytle1985introduction,PhysRevLett.66.3245,mukamel1995principles}. In Josephson superconducting circuits, pump-probe measurements are exploited for mode tuning via the AC stark shift~\cite{chu2017quantum,Lupke2022parity} and to engineer level attraction between signal and idler~\cite{Fani_2021}. In nanomechanics, it is used to investigate the coupling between vibrational modes~\cite{Westra_2010,Eichler_2012,Antoni2012Nonlinear}, study isolating electronic states in carbon nanotubes~\cite{khivrich2019nanomechanical}, and quantify the phase noise squeezing of a resonator~\cite{Ochs_2021}. In all these experiments, the probe amplitude is kept small to only explore linear perturbations of the system around the dominant pump-driven physics~\cite{khitrova1988theory}. This condition separates the impact of the pump tone that sets the stationary state from the probe that induces small fluctuations, making the method versatile and simple to interpret.

While useful for linear systems, the separation of the driving tones into so-called pump and probe severely restricts applications in nonlinear systems to the domain of linear fluctuations around the stationary states of the pump. Away from this linear approximation, the combination of both tones may lead to significant effects that cannot be analyzed separately~\cite{PhysRevA.29.1973, eberly1988phase, gaizauskasCoherentTransientsPumpprobe1994, bonacic-kouteckyTheoreticalExplorationUltrafast2005,Westra_2010,Antoni2012Nonlinear,Frimmer2019,Fani_2021}. Essentially, nonlinear dynamics map to vector-flows whose topology encode the structural arrangement of attractors and separatrices~\cite{villa2024topological}. These vectors flows will dynamically evolve due to the combined tones. Examples where the interplay of the nonlinearity with \textit{multiple} driving tones leads to new phenomena are strong optomechanical squeezing~\cite{kronwald2013arbitrarily,zhang2022mechanical}, pulse-width modulation~\cite{houri2019pulse}, parametric symmetry breaking~\cite{leuch_2016}, and routes into chaos~\cite{ide1989bifurcation,belogortsev1992bifurcations,strogatzNonlinearDynamicsChaos2007,houri2020chaos,madiot2021bichromatic,defoort2021dynamical}. In the latter two examples, the second tone drives instabilities in the system, leading to new dynamics that are challenging to describe analytically~\cite{strogatzNonlinearDynamicsChaos2007,kosata2022harmonicbalance,delpino2023limitcycles}. In such a setting, the timescale of the beating between the drives relative to relaxation times in the system plays a crucial role in the resulting dynamics.

\begin{figure}[t!]
\center
\includegraphics[scale=1]{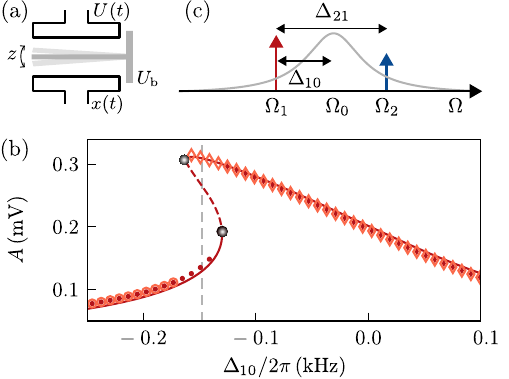}
\caption{Experimental setup. (a)~Simplified sketch of the tuning fork resonator (light grey) vibrating with displacement $z(t)$ in response to a voltage signal $U(t)$. The capacitive coupling to the lower electrode converts the vibration into an electrical signal $x(t)$. The tuning fork is biased with a voltage $U_\mathrm{b} = \SI{32}{\volt}$. (b)~Measured amplitude response $A$ to a single driving tone $U=\SI{140}{m\volt}$ swept from low to high (high to low) frequency $\Omega_1$ as filled (empty) blue{symbols}, with $\Delta_{10} = \Omega_1 - \Omega_0$. Disks and diamonds represent the lower and upper stable solution, respectively. The red solid (dashed) lines represent the analytical stationary stable (unstable) solutions, cf. Eqs.~\eqref{eq:duffing_2_tone_slow_flow}. Grey spheres highlight the positions of the bifurcation points, cf. Eq.~\eqref{eq:duffing_bif}, and the grey vertical line corresponds to the tone frequency $\Delta_{10}/2\pi=\SI{-148}{\hertz}$ used in Fig.~\ref{f:fig2}. (c)~In Figs. \ref{f:fig2} to \ref{f:fig4}, we use a second tone at frequency $\Omega_2$, detuned from the first tone by $\Delta_{21} = \Omega_2 - \Omega_1$.}
\label{f:fig1}
\end{figure}

In this work, we explore the interplay between vector flow and dynamical phase transitions in a single Duffing (Kerr) resonator subject to a combination of two near-resonant drive tones. We measure the response of the resonator by varying the relative strength and detuning of the one of the tones. Under particular conditions, the system response to the two drive tones is characterized by periodic orbits in phase space around the stationary states of the nonlinear resonator. These orbits arise from an interplay between the detuning between the two drives, multistability, and dissipation. We find two separate regimes of orbits, originating from either slow or fast changes in the vector flow relative to the decay time. We explain both regimes with a modified model of the Duffing resonator.
Our results thus highlight a key distinction in multi-tone measurements, and lay the groundwork for future explorations of nonlinear networks. Such networks have important implications for society as model systems to study tipping points in early-warning theories~\cite{Kuehn_2022, Ragone_2017, Bury_2021, strogatzNonlinearDynamicsChaos2007} that pertain, for example, to climate models, avalanches, and economy.

\section{Experimental results}
Our experimental setup consists of a double ended tuning fork resonator, made out of single-crystal silicon, which is capacitively coupled to electrodes fabricated next to it \cite{Agarwal_2008}, see Fig.~\ref{f:fig1}(a) and \cite{supmat} for additional details. We employ the lowest mechanical mode, whose resonance frequency $\Omega_0$ and Duffing nonlinearity $\beta$ is tuned by applying a bias voltage $U_\mathrm{b}$. An oscillating voltage $U(t)$ results in a drive. The resulting displacement $z(t)$ is converted into an oscillating electrical voltage $x(t)$. For simplicity, we treat $x$ as our effective degree of freedom and express all parameters accordingly as those of an effective electrical resonator. Under a strong drive, the response of the system is distinctly nonlinear and can be described by the equation of motion:
\begin{equation}\label{eq:EOM}
    \ddot{x} + \Omega_0^2 x + \Gamma \dot{x} + \beta x^3 =F(t)\,,
\end{equation}
with $\Gamma$ the damping rate, $F(t)= U(t)\times K$ the overall applied drive in units of \SI{}{\volt\per\second\squared} with a conversion factor $K\approx\SI{1e7}{\per\second\squared}$, and $\beta$ the negative (softening) Duffing coefficient. Equation~\ref{eq:EOM} does not contain a term proportional to $x^2$, as this merely leads to a modified effective value of $\beta$, see Chapter 2 in Ref.~\cite{eichler2023classical}. In our treatment, this modification is included in the usual calibration of $\beta$ and the small effect of pulling due to $\langle F^2\rangle$ is ignored. From calibration measurements, we obtain the parameters $\Omega_0/2\pi =1.109 ~\mathrm{MHz}$, $\Gamma/2\pi = 110\,\mathrm{Hz}$, and $\beta = -1.89\times 10^{17}\,\mathrm{V}^{-2}\mathrm{s}^{-2}$~\cite{supmat}.

To characterize our system, we first measure the response of the resonator to a single drive $F(t)= F_1 \cos{(\Omega_1 t+\theta_1)}$ with frequency $\Omega_1\approx\Omega_0$, fixed phase $\theta_1$, and strength $F_1/K=\SI{140}{\milli\volt}$.
We generate the oscillating drive and collect data with a lock-in amplifier \cite{ZI} set to a local oscillator frequency $\Omega_1$, enabling the down-conversion to extract the in-phase ($X$) and out-of-phase ($Y$) quadratures defined by $x(t) = X(t)\cos(\Omega_1 t) - Y(t)\sin(\Omega_1 t)$. In Fig.~\ref{f:fig1}(b), we show the measured amplitude response at frequency $\Omega_1$, $A(t)=\sqrt{X(t)^2+Y(t)^2}$, for sweeps from low to high (high to low) frequencies as filled (empty) dots. In each sweep, we observe a jump between high and low amplitudes, as expected in a Duffing resonator, reflecting transitions between stationary oscillating solutions at the driving frequency $\Omega_1$. The jump position is hysteretic and is influenced by initial conditions and sweep direction. 

\begin{figure}[t]
\center
\includegraphics[scale=1]{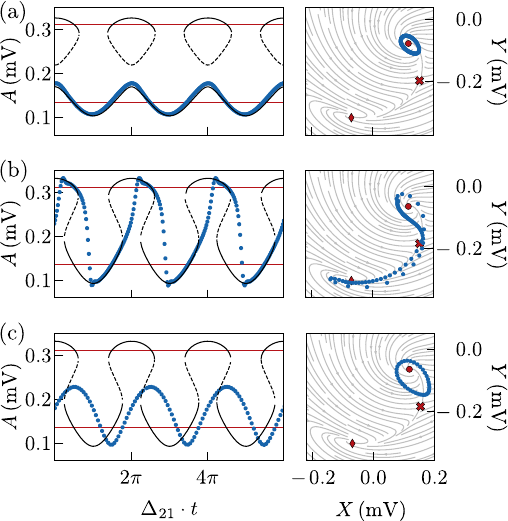}
\caption{System response to a two-tone drive. Measured amplitude $A(t)$ over time (left panel) and trajectory in phase space (right panel) for (a)~$h\equiv F_2/F_1=~0.14$ and $\Delta_{21}/2\pi=\SI{10}{\hertz}$,  (b)~$h=~0.21$ and $\Delta_{21}/2\pi=\SI{10}{\hertz}$, and (c)~$h=~0.21$ and $\Delta_{21}/2\pi=\SI{90}{\hertz}$. We initialize the system in the low-amplitude state with a fixed voltage $F_1/K=\SI{140}{m\volt}$ and $\Delta_{10}/2\pi\approx\SI{-148}{\hertz}$, cf. Fig.~\ref{f:fig1}(b). Blue points are measured data. On the left, the dark solid (dashed) line represents the calculated time-dependent stable (unstable) stationary solutions, while the red lines indicate the analytical single-tone stationary solutions of the system for $h=0$ in Eq.~\eqref{eq:duffing_2_tone_slow_flow}. On the right, the disk (diamond) represents the low (high) amplitude stable state, corresponding to the red lines on the left. The cross marks the unstable state, and grey lines show the vector flow arising from Eq.~\eqref{eq:duffing_2_tone_slow_flow} for $h=0$. 
}
\label{f:fig2}
\end{figure}

Next, we add a second weak probe tone, such that $F(t)= \sum_{m=1,2} F_m \cos{(\Omega_m t+\theta_m)}$. We initialize the system in the low-amplitude stationary solution in Fig.~\ref{f:fig1}(b) at a detuning $\Delta_{10} = \Omega_1-\Omega_0$, cf. Fig.~\ref{f:fig1}(c). The second tone has an amplitude $F_2 = h F_1$, and a frequency $\Omega_2$ detuned from $\Omega_1$ by $\Delta_{21}=\Omega_2 - \Omega_1$. For $\Delta_{10}/2\pi = \SI{-148}{\hertz}$, $h= 0.14$, and $\Delta_{21}\approx 0.09\Gamma$, the response at $\Omega_1$ shows small oscillations around the initial solution (red line and symbols in Fig.~\ref{f:fig2}) with frequency $\Delta_{21}$, see left panel in Fig.~\ref{f:fig2}(a). In the frame rotating at $\Omega_1$, spanned by $X$ and $Y$, this oscillation forms a closed loop around the initial state, see right panel. 
Increasing the probe amplitude to $h = 0.21$, we observe a striking change in the response: as shown in Fig.~\ref{f:fig2}(b), the system exhibits large amplitude variations, with $X$ and $Y$ tracing an eight-shaped trajectory which circles the two stable solutions the system has when driven only by $F_1$ ($h=0$). Increasing the detuning $\Delta_{21}$ while keeping $h=0.21$ alters the response once more, see Fig.~\ref{f:fig2}(c). Here, the system's quadratures oscillate again around the low-amplitude solution. However, as is discussed below, this case differs from the similar phenomenon in Fig.~\ref{f:fig2}(a), marking a third regime with unique features.
\section{Effective model}

To better understand the differences between the three cases, we write Eq.~\eqref{eq:EOM} in the rotating frame measured by the lock-in amplifier at the frequency of the first drive. Assuming the quadratures \(X(t), Y(t)\) to vary slowly relative to the drive frequency \(\Omega_1\), we average them over the oscillation period $2\pi/\Omega_1$~\cite{kosata2022harmonicbalance, eichler2023classical}. This yields
\begin{equation}
\resizebox{0.905\columnwidth}{!}{$\displaystyle{\begin{aligned}\label{eq:duffing_2_tone_slow_flow}
    \dot{X}=&-\Gamma\frac{X}{2}-\frac{Y}{2}\left(\frac{3\beta}{4\Omega_1}A^{2}+\frac{\Omega_{0}^{2}-\Omega_{1}^{2}}{\Omega_1}\right)+\frac{\mathrm{Im}(f(t))}{2\Omega_1},\\
    \dot{Y}=&-\Gamma\frac{Y}{2}+\frac{X}{2}\left(\frac{3\beta}{4\Omega_1}A^{2}+\frac{\Omega_{0}^{2}-\Omega_{1}^{2}}{\Omega_1}\right)-\frac{\mathrm{Re}(f(t))}{2\Omega_1},
\end{aligned}}$}
\end{equation}
with $f(t) = F_1e^{i\theta_1}+F_2e^{i(\varphi(t)+\theta_2)}$ and $\varphi(t)=\Delta_{21}t$. For Eq.~\eqref{eq:duffing_2_tone_slow_flow} to hold, $|\beta| A^{2}/\left(\Omega_{1}^{2}\right)$, $\left|\Omega_{1}^{2}-\Omega_{0}^{2}\right|/\Omega_{1}^{2}$, and $\sqrt{|\beta| F_{m}^{2}/\Omega_{1}^{6}}$ must be $\ll 1$~\cite{eichler2023classical}; conditions always fulfilled by our experimental parameters. Equations~\eqref{eq:duffing_2_tone_slow_flow} are valid when the variation of $f(t)$, driven by the detuned second drive and its relative phase $\varphi(t)$, is sufficiently slow. The averaging is carried out to second order, introducing a small correction from residual quadratic nonlinearities (additional terms $\propto x^2$ in Eq.~\ref{eq:EOM}), that renormalizes the effective Duffing nonlinearity~\cite{eichler2023classical}; this corrected $\beta$ is the one extracted from the experiment~\cite{supmat}.

Intuitively, we can understand the main effect of the two detuned drives as a single amplitude-modulated drive with
\begin{equation}\label{eq:U_eff}
     |f(t)| = F_1\sqrt{1+h^2+2h\cos(\theta_1-\theta_2-\varphi(t))}\,.
\end{equation}
Within each period $1/\Delta_{12}$, the resonator experiences a drive amplitude ranging from $F_1 - F_2$ (drives are completely out of phase) to $F_1 + F_2$ (drives are in phase). For an effective drive $f$ at a single point in time, we can  set $\dot{X} = \dot{Y} = 0$ in Eqs.~\eqref{eq:duffing_2_tone_slow_flow} and obtain either one or two stable stationary solutions at frequency $\Omega_1$~\cite{kosata2022harmonicbalance,eichler2023classical}. For $h=0$, this yields the stationary amplitudes and bifurcations of the single-tone Duffing resonator, see Fig.~\ref{f:fig1}(b)~\cite{supmat}. The bifurcation points occur when
\begin{equation}\label{eq:duffing_bif}
   \resizebox{0.905\columnwidth}{!}{$\displaystyle{|f|^{2}=\frac{8}{81\beta}\left[\left(\tilde\Delta_{01}^{4}-3\Gamma^{2}\Omega_{1}^{2}\right)^{\frac{3}{2}}\pm\tilde\Delta_{01}^{2}\left(\tilde\Delta_{01}^{4}+9\Gamma^{2}\Omega_{1}^{2}\right)\right],}$}
\end{equation}
with $\tilde\Delta_{01}=\sqrt{\Omega_{0}^{2}-\Omega_{1}^{2}}$ ~\cite{supmat}. The effective drive $f$ therefore sets the position of the two bifurcation frequencies. Since the drive is amplitude modulated, the detunings $\Delta_{10}$ at which the bifurcation points occur are also modulated, cf. Eq.~\eqref{eq:duffing_bif}. If the modulation is strong enough, the number of stable solutions at $\Omega_1$ changes between 1 and 2 periodically.

\begin{figure}[t]
\center
\includegraphics[scale = 1]{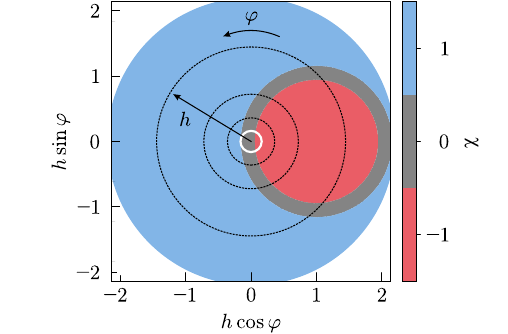}\caption{Calculated topological phase diagram for the two-tone driven Duffing resonator. We use $F_1/K=\SI{140}{m\volt}$ and $\Delta_{10}/2\pi\approx \SI{-148}{\hertz}$ as in~Fig.~\ref{f:fig2}. Depending on the amplitude $h=F_2/F_1$ and the phase $\varphi$ between the drives [cf. Eq.~\eqref{eq:U_eff}], the system can have a net chirality $\chi=1$, $0$, or $-1$, corresponding to distinct topological phases that offer only the high-amplitude solution (monostable), both solutions (bistable), or only the low-amplitude solution (monostable), respectively. The white ring represents the critical value of $h$ to enter the high-amplitude monostable region.}
\label{f:fig3}
\end{figure}

Note that Eq.~\eqref{eq:duffing_2_tone_slow_flow} defines a 2D vector flow, shown as grey lines in the phase space graphs in Figs.~\ref{f:fig2}. Different solutions of the system are characterized by different flow chiralities, where we assign the value $1$ ($-1$) to a clockwise (counter-clockwise) chirality, highlighting whether the dynamics is advanced or delayed relative to the driving tone. In the Duffing resonator with $\beta<0$, the upper (lower) solution has a chirality of $1$ ($-1$)~\cite{soriente2020distinguishing,Soriente2021,dumont2024hamiltonian}. We can solve for all stable stationary solutions $i\in\{1,2\}$ as a function of $h\cos\varphi$ and $h\sin\varphi$, and at each point sum over their chiralities $\chi_i$, see Fig.~\ref{f:fig3}. This yields a net chirality $\chi=\sum_i\chi_i$. When the low-and high-amplitude solutions coexist, $\chi=0$. The net chirality is a sufficient topological index for our system~\cite{villa2024topological}. To summarize, structural changes in the vector flow manifest as changes in the number of solutions and their local flow chirality, indicating topological phase transitions.

\begin{figure*}[t]
\center
\includegraphics[scale = 1]{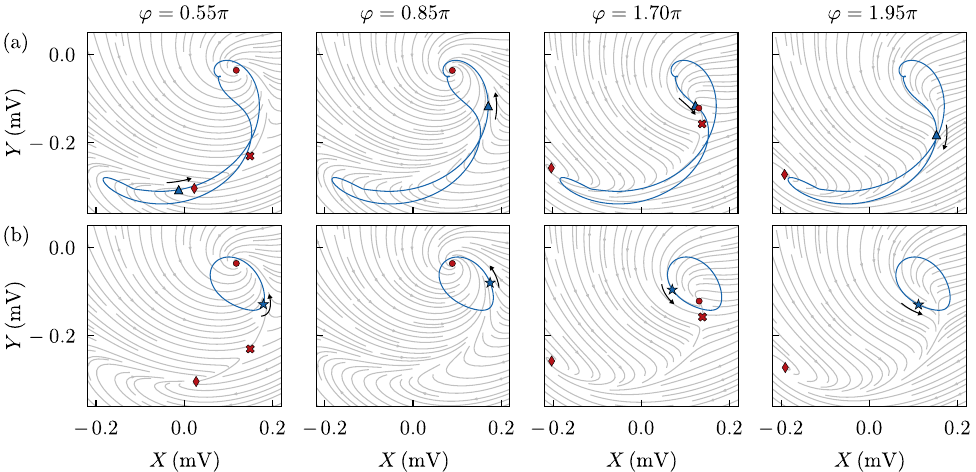}
\caption{Simulated qualitative difference between the slow and fast regime. (a) From left to right we show snapshots at various $\varphi$ (top labels), during a single modulation cycle of the slow ($\Delta_{21}/2\pi=10\,\mathrm{Hz}$) regime. The instantaneous solution (blue triangle) initially continuously follows the high amplitude solution (red diamond), jumps to the low-amplitude solution (red disk), continuously follows the low-amplitude solution, and then jumps to the high-amplitude solution. (b) Same snapshots in the fast ($\Delta_{21}/2\pi=90\,\mathrm{Hz}$) regime. Here, in contrast, the instantaneous solution (blue star) remains confined to a trajectory around the low-amplitude solution. Parameters are the same as in the corresponding experiments in Fig.~\ref{f:fig2}, with $\varphi=\Delta_{21}t$. The analytically calculated positions of the two stable (red disk and diamond) and the unstable solution (red cross), as well as the instantaneous system position, are shown together with a numerical simulation of the entire trajectory (blue line). Grey lines represent the stationary vector flow calculated from Eqs.~\eqref{eq:duffing_2_tone_slow_flow} assuming an effective instantaneous force $f$ [Eq.~\eqref{eq:U_eff}] with $h=0.21$ and a different phase $\varphi$ in each panel (top labels).}
\label{f:fig4}
\end{figure*}

We return now to our two-tone dynamical system. For a fixed second drive with relative amplitude $h>0$, the system traverses a circular path in the stationary phase diagram [Fig.~\ref{f:fig3}] as its phase $\varphi$ evolves over time. The path can traverse regions with different $\chi$ and may induce  \textit{dynamical phase transitions}~\cite{Heyl2013} between low- and high-amplitude solutions. Nonetheless, the system’s ability to undergo a dynamical phase transition and reach the vicinity of a new solution depends on the available time before the solution vanishes again. To determine which solutions are reached over time in the experiment, we track the stationary amplitudes along each circular trajectory in Fig.~\ref{f:fig3} and overlay them with the observed $A(t)$ in the left panels of Fig.~\ref{f:fig2}. 
When the drive modulation is relatively small ($h=0.14$), the low-amplitude solution never loses stability, and the system remains trapped near the initial single-tone solution, see Fig.~\ref{f:fig2}(a). In contrast, the case with an increased probe strength ($h=0.21$) in Fig.~\ref{f:fig2}(b) corresponds to a trajectory in Fig.~\ref{f:fig3} that enters both the low-amplitude monostable region (red) and the high-amplitude monostable region (blue). The system follows stable solutions until they vanish at a bifurcation point, forcing a periodic jump between low- and high-amplitude Duffing solutions. This corresponds to repeated dynamical phase transition imposed periodically by $\varphi(t)$. As the system has enough time to probe the change in topology, we term this regime `slow'.

The slow regime ($\Delta_{2,1}<\Gamma$), which we explore in Fig.~\ref{f:fig2}(b), was also studied in Ref.~\cite{houri2019pulse}. It can be compared to an hourglass where all the sand collects on the lower side before being turned. Such phenomena are studied in a broad range of fields, ranging from bursting oscillations observed in nonlinear coupled systems~\cite{zhang2023bursting} to tipping points of avalanches~\cite{strogatzNonlinearDynamicsChaos2007,ma2024relaxation}. By contrast, in Fig.~\ref{f:fig2}(c), we used $\Delta_{21} \approx \Gamma$. Even though the stationary solutions and chiralities are identical to the ones in Fig.~\ref{f:fig2}(b) and we follow the same path in Fig.~\ref{f:fig3}, the system cannot probe the changing vector-flow topology quickly enough. The jumps are thus never completed, and the system orbits around the initial single-tone solution. To distinguish from the slow case above, we term this regime `fast'. The fast case contrasts with Fig.~\ref{f:fig2}(a), where the system can follow a single stationary state continuously because that solution remains available during the entire period. 

To emphasize the difference between the slow regime in Fig.~\ref{f:fig2}(b) and the fast regime in Fig.~\ref{f:fig2}(c), Fig.~\ref{f:fig4} presents snapshots of the accumulated system's trajectory at different times during a single period, shown alongside the instantaneous vector flow. The vector flow lines are calculated using Eqs.~\eqref{eq:duffing_2_tone_slow_flow}, with the same parameters as in Fig.~\ref{f:fig2}(b) and Fig.~\ref{f:fig2}(c). A numerical simulation of the slow regime is shown in Fig.~\ref{f:fig4}(a). Like in the experiment in Fig.~\ref{f:fig2}(b), we observe the characteristic eight-shaped trajectory due to the opposite chirality of the flow lines around the two solutions. In the fast regime in Fig.~\ref{f:fig4}(b), instead, the simulated trajectory is strikingly different, even though the global topology of the flow lines is identical. Here, the system is trailing (but never quite catching up with) the low-amplitude solution or the changing topology of the flow lines. The resulting small loop strongly resembles the experimental observation in Fig.~\ref{f:fig2}(c). The time-dependent flow analysis thus reveals stark contrast between the slow and fast regimes, driven by the interplay of modulation timescales and dissipation, and traceable to the system’s capacity—or lack thereof—to adapt to changing conditions. Interestingly, a systematic study of the system response to various $\Delta_{12}$ and $h$ reveals a non-trivial transition between $fast$ and slow regime, see Section III of \cite{supmat}. We leave the analytical description of such interplay to future work.

\section{Conclusion and Outlook}
This work exemplifies the competition between an amplitude-modulated drive and resonator damping in a nonlinear oscillator. In particular, we highlight the transition from a slow regime, where the system adheres to the underlying topology of the phase-space dynamics and undergoes dynamical phase transitions, to a fast regime where this is no longer true. While starting from a similar situation as previous works~\cite{ide1989bifurcation,belogortsev1992bifurcations,strogatzNonlinearDynamicsChaos2007,houri2020chaos,madiot2021bichromatic,defoort2021dynamical}, we thus emphasize a fundamental aspect that was hitherto not well addressed: \textit{why} does the system select a particular trajectory in response to multi-tone drives, and how can it be controlled? The insights from this example are broadly applicable to other nonlinear systems, particularly to networks of nonlinear resonators, where the coupling terms of the individual network constituents act as multiple drive tones that can induce many-body oscillation phases~\cite{Bello2019,Matheny2019,Heugel2019,zaletel2023colloquium}. Characterizing these phases in terms of their time-dependent phase-space topology~\cite{villa2024topological} and departures therefrom will provide a robust framework for sensing applications~\cite{houri2019pulse}, computer vision~\cite{Hamerly2019}, and kinematic synthesis in complex machinery~\cite{angeles2022kinematics}, as well as further our understanding of water wave dynamics~\cite{simonelli1989surface,apffel2024experimental} and combustion engines~\cite{stoychev2024nonlinear}, as well as early-warning models~\cite{strogatzNonlinearDynamicsChaos2007} employed in the context of avalanches, earthquakes, and solar flares~\cite{ma2024relaxation}. 

\section*{Acknowledgments}
The authors acknowledge Christian Marti and Sebasti\'an Guerrero for help building the measurement setup. We thank
Nicholas E. Bousse and T. W. Kenny for providing the MEMS device, and Joel Moser for inspiring discussions. V. D. acknowledges support from the ETH Zurich Postdoctoral Fellowship Grant No.~23-1 FEL-023. O. Z. acknowledges funding from the Deutsche Forschungsgemeinschaft (DFG) via project number 449653034 and through SFB1432. A. E. and O. Z. acknowledge financial support from the Swiss National Science Foundation (SNSF) through the Sinergia Grant No.~CRSII5\_206008/1.

\appendix
\begin{widetext}
\section*{Appendices}
\section{Device and Characterization}\label{app:char}
The device we use, fabricated by T. Kenny's group in Stanford \cite{Agarwal_2008}, is a highly-doped single-crystal silicon microelectromechanical system (MEMS) with a p-type (Boron) concentration of $1.5\times10^{20}\si{cm}^{-3}$. The MEMS device has the shape of a double-ended tuning fork with branches $200 \si{\mu m}$ long and $6 \si{\mu m}$ thick. The resonator is encapsulated through an epi-seal process at a pressure of $~10^{-1}\si{mbar}$ \cite{Yang2016}. As discussed in the main text, the vibration of the resonator $z$ is converted into an electric signal $x$ through the electrodes fabricated next to the resonator. A schematic of the device and of the electrical connections is shown in Fig. \ref{fig:suppl_scheme}. A finete element simulation of the device motion between the electrodes can also be found in \cite{Polunin_2016} All the numerical simulations are performed using individually measured parameters. Here we describe the characterization procedure used to extract them.

\begin{figure}[h]
    \centering
    \includegraphics[scale = 1]{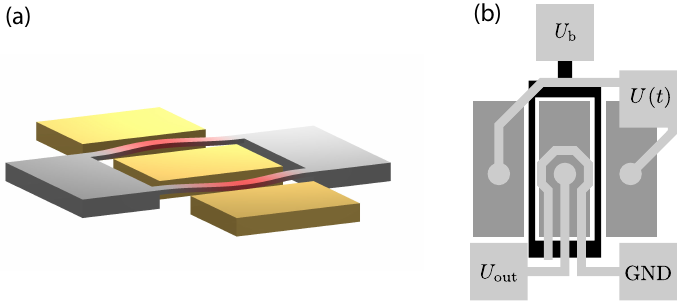}
    \caption{MEMS device. (a) Schematic representation of the double-ended tuning fork (grey) vibrating between the two driving electrodes (gold). (b) Schematic of the MEMS electrical connections. In black we draw the double-ended tuning fork, in dark grey the electrodes, in light grey the pad and electrical connections. $U_\mathrm{b}$ is the bias voltage, $U(t)$ the driving voltage and $U_\mathrm{out}$ the readout voltage measured with respect to ground (GND). The schematic is reproduced from~\cite{Margiani2023}.}
    \label{fig:suppl_scheme}
\end{figure}

By applying an oscillating voltage $U(t)$, we can apply a driving `force' $F(t)=U(t)\times K$ in units of $\si{V}\si{s}^{-2}$, where $K$ is a conversion factor in units of $\si{s}^{-2}$. In the limit of weak drive, the resonator is governed by the equation of motion 
\begin{equation}\label{eq:EOM_lin}
\ddot{x}(t)+\Omega_0^2x(t)+\Gamma\dot{x}=F(t),
\end{equation}
with $\Omega_0$ the resonance frequency, $\Gamma$ the damping rate and $F(t)= F_1\cos(\Omega_1t+\theta_1)$. By solving Eq. \eqref{eq:EOM_lin} in the frequency domain, we find the resonator's response to an external drive:
\begin{equation}\label{eq:linresponse}
x = \frac{F_1e^{i\theta_1}}{\Omega_0^2-\Omega_1^2-i\Gamma\Omega_1}.
\end{equation}
The vibration $x$ contains both the amplitude and the phase response, described by the relations $A=|x|$ for the amplitude and $\phi=\arg(x)$ for the phase.

To characterize the linear mechanical parameters, we sweep the driving frequency $\Omega_1$ across the resonance. We use a Zurich Instruments MFLI lock-in amplifier with a local oscillator frequency $\Omega_1$ to generate the driving tone and measure the two quadratures $X(t)$ and $Y(t)$ demodulated at $\Omega_1$, which satisfy the relation $x= X(t)\cos(\Omega_1t)-Y(t)\sin(\Omega_1t)$. For each driving frequency, we can extract the amplitude $A=\sqrt{X^2+Y^2}$ and phase $\phi=\mathrm{atan2}{(-Y/X)}$ of the resonator as a function of frequency, red dots in Fig.~\ref{fig:suppl_linfit}. 
\begin{figure}[h]
    \centering
    \includegraphics[scale = 1]{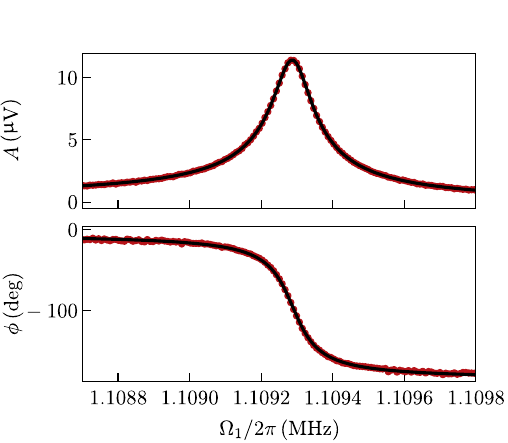}
    \caption{Linear response of the resonator to an external weak force $F_1/K= 5\si{mV}$. The red dots are the experimental data, the black solid lines corresponds to the fitted model $A=|x|$ and $\phi=\arg(x)$.}
    \label{fig:suppl_linfit}
\end{figure}
We then perform a fit of the amplitude $A$ and of the phase $\phi$, black lines in Fig.~\ref{fig:suppl_linfit}. From the two fits, we extract the parameters $\theta_1 = 9(1)$~degrees, $\Omega_0/2\pi=1.10928791(2)~\si{MHz}$ and $\Gamma/2\pi=108.25(4)~\si{Hz}$, leading to $Q=\Omega_0/\Gamma=10246(4)$. From the amplitude fit, we also extract the conversion factor $K=1.092(2)\,\si{s}^{-2}$. Note that the uncertainties reported are only statistical errors from the fit. The real errors on the parameters is larger due to drifts in the resonance frequency.

In a next step, we move to extract the nonlinear parameter $\beta$. To do so, we again measure the response of the resonator as we sweep the driving frequency across the resonance after increasing the driving voltage to $F_1/K=140~\si{mV}$. At this voltage, our system behaves as a Duffing resonator featuring an amplitude-dependent resonance \cite{Bachtold_2022}
\begin{equation}\label{eq:backbone}
    \delta \Omega_0 =  \frac{3\beta}{8\Omega_0} A^2,
\end{equation}  
and hysteresis in the frequency response. Thus, we repeat the same measurement twice, first sweeping the frequency $\Omega_1$ from high to low frequencies [Fig.~\ref{fig:suppl_nlfit}(a)] and from low to high frequencies [Fig.~\ref{fig:suppl_nlfit}(b)].
\begin{figure}[h]
    \center
    \includegraphics[scale =1]{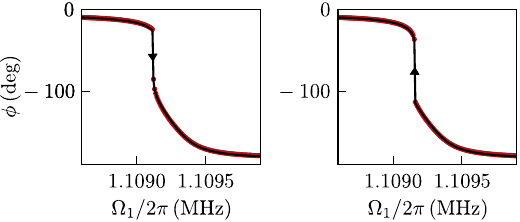}
    \caption{Nonlinear response of the resonator to an external force $F_1/K= 140\si{mV}$. Stationary phase due to an external driving force with frequency $\Omega_1$, swept from (left) high to low and (right) low to high frequencies, as shown in arrows. The black line is a two-variable fit obtained  using Eq. \eqref{eq:nlphase} as model and the stationary amplitude measured simultaneously as second variable. See text for details.}
    \label{fig:suppl_nlfit}
\end{figure}

By substituting $\Omega_0$ with $\Omega_0 + \delta\Omega_0$ into Eq.~\eqref{eq:linresponse}, we find the nonlinear mechanical susceptibility
\begin{equation}\label{eq:nlresponse}
    x_\mathrm{nl}=\frac{F_1e^{i\phi_1}}{(\Omega_0^2-\Omega_1^2+\frac{3}{4}\beta A^2)+i\Gamma \Omega_1},
\end{equation}
which we can use to write the amplitude and phase response. The presence of the nonlinear term makes the fitting procedure challenging. Instead of performing two independent fits, we use the simultaneously measured amplitude and phase data to perform a two-variable fit of the phase response
\begin{equation}\label{eq:nlphase}
\phi = \mathrm{atan2}\left(\frac{-\Omega_1\Gamma}{\Omega_0^2-\Omega_1^2+\frac{3}{4}\beta A^2}\right),
\end{equation}
where the measured amplitude and the driving frequency serve as the two variables. From the fit, we extract the Duffing parameter $\beta=-1.8875(5)\times 10^{17}~\mathrm{V^{-2}s^{-2}}$. 
\section{Trajectories in phase space}
To improve the visibility of the phase space trajectories reported in Fig. 2, we show here the same plots with a larger size (Fig. \ref{fig:suppl_trajectories}).

\begin{figure*}[h]
    \center
    \includegraphics[scale =1]{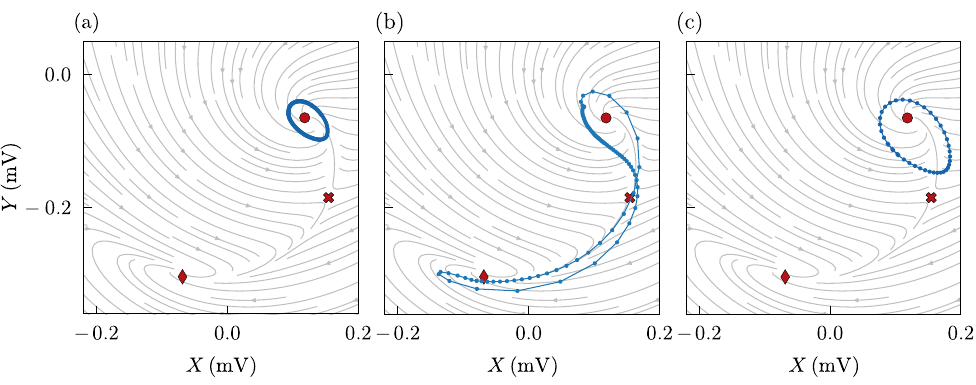}
    \caption{Zoom-in of the trajectories in phase space. The data shown in this figure are the same shown in Fig. 2 of the main text with (a)~$h\equiv F_2/F_1=~0.14$ and $\Delta_{21}/2\pi=10~\si{Hz}$,  (b)~$h=~0.21$ and $\Delta_{21}/2\pi=10~\si{Hz}$, and (c)~$h=~0.21$ and $\Delta_{21}/2\pi=90~\si{Hz}$. The blue points are data and we plotted them connected by a solid line to highlight the shape of the trajectories. The red disk (diamond) represents the lower (upper) stable solutionsm the red cross the unstable solution. The grey lines show the vector flow arising from Eq. (2) in the main text for $h=0$}
    \label{fig:suppl_trajectories}
\end{figure*}

\section{Effect of driving strength and detuning}
The transition from a slow regime to a fast one, observed in Fig.~2 and simulated in Fig.~4, is governed by the competition between the amplitude-modulated drive and the resonator damping. To increase our understanding of the system response to different driving strength and detuning, we systematically probe the response of the system as a function of $\Delta_{21}$ and $h$ for fixed
$F_1$ and $\Omega_1$. For each combination, we initialized the system in the lower solution and extract the average value of the amplitude $\bar{A}$ from an interval of $5~\si{s}$. In Fig.~\ref{fig:suppl_phasediag}, we show the complete set of $\bar{A}$ for different values of $\Delta_{21}$ and $h$. We distinguish two main regions: one region has low $\bar{A}$. Here, the second tone merely modulates the vibrational amplitude without inducing jumps, cf. Fig. 2(a) and Fig. 2(c). A second region has high $\bar{A}$ associated with jumping between the high and low-amplitude solutions, cf. Fig. 2(b).

The overall trend of the experiment indicates that larger detuning $\Delta_{21}$, corresponding to faster
modulation in the frame rotating at $\Omega_1$, requires higher drives $h$ to induce jumps. A thorough study of this behavior is left for an ongoing investigation.

\begin{figure*}[h]
    \center
    \includegraphics[scale =1]{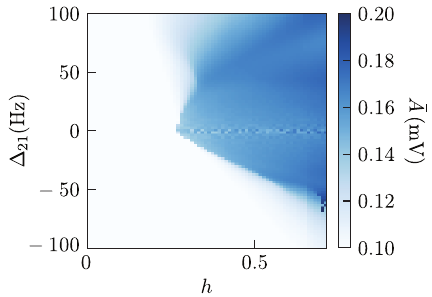}
    \caption{Average amplitude $\bar{A}$ for different detuning $\Delta_{21}$ and relative driving strength $h$. Low $\bar{A}$ indicates that the system remains close to the lower stable solution, while high $\bar{A}$ indicates jumping between two solutions. To collect data we used $F_1/K = 140~\si{mV}$ and  $\Delta_{10}/2\pi\approx \SI{-158}{\hertz}$.}
    \label{fig:suppl_phasediag}
\end{figure*}

\section{Effective Model: Derivation and Simulation}

\subsection{Effective slow-flow model}\label{subsec:effective_slow_flow}
The dynamics of a Duffing oscillator driven by a single periodic drive of strength $F_1$ is governed by 
\begin{equation}\label{Eq_Duffing_SinglePump}
   \ddot{x}+ \Gamma \dot{x}+ \Omega_0^2 x+\beta x^3={F_1} \cos \left(\Omega_1 t+\theta_1\right)\,,
\end{equation}
where $\Omega_1$ is the pump frequency, and $\theta_1$ is the initial phase between the pump and the system. We apply the Van der Pol transformation $x(t) = X(t) \cos \left(\Omega_1 t\right) - Y(t) \sin \left(\Omega_1 t\right)$ \cite{eichler2023classical} to describe the dynamics in a frame rotating at $\Omega_1$, the frequency of the drive. The in-phase and quadrature components, $X(t)$ and $Y(t)$ respectively, describe the exact dynamics of the system with terms in the resulting equation oscillating at frequencies $\Omega_1$ and $3\Omega_1$. However, this system is not amenable to a closed-form analytical solution. For a perturbative solution, we distinguish between the timescales governing the evolution of the quadratures and the rotating frame via the slow-flow approximation \cite{rand2012lecture}. It entails that the quadratures $X(T), Y(T)$ depend on \textit{slow time} \cite{rand2012lecture} $T\gg \frac{2\pi}{\Omega_1}$ and evolve slowly, allowing us to ignore higher-order derivatives $\ddot{X}, \ddot{Y}$. Averaging over one cycle $t\mapsto t+T$, we obtain the so-called slow-flow equations,
\begin{equation}
{\begin{aligned}\label{eqapp:duffing_2_tone_slow_flow}
   \dot{X}=&-\Gamma\frac{X}{2}-\frac{Y}{2}\left(\frac{3\beta}{4\Omega_1}A^{2}+\frac{\Omega_{0}^{2}-\Omega_{1}^{2}}{\Omega_1}\right)+\frac{F_1 \sin \theta}{2\Omega_1},\\
   \dot{Y}=&-\Gamma\frac{Y}{2}+\frac{X}{2}\left(\frac{3\beta}{4\Omega_1}A^{2}+\frac{\Omega_{0}^{2}-\Omega_{1}^{2}}{\Omega_1}\right)-\frac{F_1 \cos \theta}{2\Omega_1}.
\end{aligned}}
\end{equation}
This is equivalent to isolating the terms oscillating at $\Omega_1$ in the transformed equation of motion and ignoring the higher order derivatives according to the harmonic balance method \cite{krack_2019}.

For the stationary state solution, we solve the above equations for $\dot{X}=0, \dot{Y}=0$. Squaring and adding the resultant equations, we obtain
\begin{equation}\label{eq:polynomial}
    A^2\left( \frac{\Gamma^2}{4} + \left( \frac{3\beta A^2 }{4\Omega_1} + \frac{\Omega_1^2 - \Omega_0^2}{\Omega_1} \right)^2 \right) = \frac{F_1^2}{4\Omega_1^2}.
\end{equation}
This is a cubic polynomial in $A^2$ and its real roots give the different stationary state amplitudes of the Duffing resonator. The discriminant of a polynomial $D$ provides a condition to determine when its roots change behaviour, such as merging or splitting, indicating critical points like bifurcations. The roots of the discriminant \( D\) separate regions where the cubic equation has three distinct real roots (\(D > 0\)), a degenerate real root (\(D = 0\)), or one real root and two complex conjugate roots (\(D < 0\)), i.e. they mark the bifurcation points.  We derive the discriminant of the polynomial in Eq. \eqref{eq:polynomial}
\begin{equation}\label{discriminant_DuffingPolynomial}
   D = 243 \beta^2 F_1^4-48 \beta F_1^2 \left(\Omega_0^2-\Omega_1^2\right)\left(9 \Gamma^2 \Omega_1^2+4 \left(\Omega_0^2-\Omega_1^2\right)^2\right) +16 \Gamma^2 \Omega_1^2 \left(\Gamma^2 \Omega_1^2+4 \left(\Omega_0^2-\Omega_1^2\right)^2\right)^2,
\end{equation}
and find its real roots to determine the critical pump strength at which a bifurcation occurs. The critical pump strengths for the low and high amplitude solution branches,  $F_{\text{lb}}$  and  $F_{\text{hb}}$, are given by:
\begin{equation} 
   F_{\text{hb(lb)}}^{2}=\frac{8}{81\beta}\left[\left(\tilde\Delta_{01}^{4}-3\Gamma^{2}\Omega_{1}^{2}\right)^{\frac{3}{2}}\pm\tilde\Delta_{01}^{2}\left(\tilde\Delta_{01}^{4}+9\Gamma^{2}\Omega_{1}^{2}\right)\right],
\end{equation}
where $\tilde\Delta_{01}=\sqrt{\Omega_{0}^{2}-\Omega_{1}^{2}}$.

In the case of a system driven by two tones, $F(t)= \sum_{m=1,2} F_m \cos{(\Omega_m t+\theta_m)}$, we note that the drive can be rewritten as
    \begin{align}\label{ForceAfterFreqDecomposition}
\sum_{m}F_{m}\cos\left(\Omega_{m}t\right)&=\cos\left(\Omega_1t\right)\left(F_{1}\cos\theta_1+\sum_{m>1}F_{m}\cos\left((\Omega_{m}-\Omega_1)t + \theta_{m}\right)\right)\nonumber\\
&-\sin\left(\Omega_1t\right)\left(F_{1}\sin\theta_1 + \sum_{m>1}F_{m}\sin\left((\Omega_{m}-\Omega_1)t + \theta_{m}\right)\right).
\end{align}

The above decomposition of the drive is equivalent to writing the different drive frequencies as $\Omega_i=\Omega_1+\Delta_{im}$, where $\Delta_{im}=\Omega_i-\Omega_m$.
Equation \eqref{ForceAfterFreqDecomposition} shows that the second drive induces amplitude modulation of the drive at $\Omega_1$ with a frequency $\Delta_{21}$. Replacing the drive terms in Eq. \eqref{ForceAfterFreqDecomposition} into Eq. \eqref{Eq_Duffing_SinglePump}, and averaging under the assumption of slow modulation to ensure slowly varying quadratures, we derive effective equations of motion akin to Eq. \eqref{eqapp:duffing_2_tone_slow_flow}: 
\begin{equation}\label{eq:duffing_two_drive_slowflow}
\begin{aligned}
\dot{X}=&-\Gamma\frac{X}{2}-\frac{Y}{2}\left(\frac{3\beta}{4\Omega_1}A^{2}+\frac{\Omega_1^{2}-\Omega_0^{2}}{\Omega_1}\right)+\frac{F_1\sin{\theta_1}+F_2 \sin (\varphi(t) + \theta_2)}{2 \Omega_1}, \\
\dot{Y}=&-\Gamma\frac{Y}{2}+\frac{X}{2}\left(\frac{3\beta}{4\Omega_1}A^{2}+\frac{\Omega_1^{2}-\Omega_0^{2}}{\Omega_1}\right)-\frac{F_1\cos{\theta_1}+F_2 \cos (\varphi(t) + \theta_2)}{2 \Omega_1},
\end{aligned}
\end{equation}
where $\varphi(t) = \Delta_{21}t$ captures the effect of the slow modulation on the drive. The quadratures have an implicit time-dependent drive through $\varphi(t)$.  

\subsection{Effect of Quadratic Nonlinearity}
A quadratic force term, corresponding to a cubic potential, can naturally arise in microelectromechanical systems, and we analyze its effect in the following. In this case, 
\eqref{Eq_Duffing_SinglePump} can be written as 
\begin{equation}
       \ddot{x}+ \Gamma \dot{x}+ \Omega_0^2 x+\beta_2x^2 + \beta_3 x^3={F_1} \cos \left(\Omega_1 t+\theta_1\right)\,,
\end{equation}
distinguishing between the quadratic nonlinearity $\beta_2$ and the cubic nonlinearity $\beta_3$ to observe the effect of the quadratic term $\propto x^2$. 
As can be directly deduced from the ansatz in \autoref{subsec:effective_slow_flow}, this term generates a second harmonic and a DC shift via a zero-frequency component. It therefore yields no contribution at the drive frequency, as the second harmonic term averages out to first averaging order~\cite{eichler2023classical}. Corrections only arise in higher-order averaging method~\cite{krylov_introduction_2016}. The second order corrections, which build on the first-order terms~\cite{krylov_introduction_2016, eichler2023classical} are computed using the methodology  implemented in HarmonicBalance.jl~\cite{HarmonicBalance.jl}. We perform this check with a single-tone drive to illustrate the effect in the simplest setting. We find:

\begin{align}\label{eq:duffing_two_drive_slowflow_2nd}
\dot{X} =&\- \frac{\Gamma X}{2}
- \frac{Y \left( \Gamma^2 \Omega_1^2 + 5 \Omega_1^4 - 6 \Omega_1^2 \Omega_0^2 + \Omega_0^4 \right)}{8 \Omega_1^3}
+ \frac{3 \beta_3 Y A^2 (2 \Omega_1^2 - \Omega_0^2)}{8 \Omega_1^3}
- \frac{5 \beta_2^2 Y A^2}{12 \Omega_1^3}
- \frac{51 \beta_3^2 Y A^4}{256 \Omega_1^3} \nonumber\\
&+ \frac{3 F_1 \beta_3}{32 \Omega_1^3} \left( \sin\theta_{1}(A^2 + 2Y^2) - 2 X Y \cos\theta_{1} \right)
+ \frac{F_{1}}{8 \Omega_{1}^3} \left(\Omega_{0}^2 \sin\theta_{1} - \Gamma \Omega_1 \cos\theta_{1} - 5 \Omega_1^2 \sin\theta_{1} \right),
\\
\dot{Y} =&\ 
- \frac{\Gamma Y}{2}
+ \frac{X \left( \Gamma^2 \Omega_1^2 + 5 \Omega_1^4 - 6 \Omega_1^2 \Omega_0^2 + \Omega_0^4 \right)}{8 \Omega_1^3}
- \frac{3 \beta_3 X A^2 (2 \Omega_1^2 - \Omega_0^2)}{8 \Omega_1^3}
+ \frac{5 \beta_2^2 X A^2}{12 \Omega_1^3}
+ \frac{51 \beta_3^2 X A^4}{256 \Omega_1^3}\nonumber \\
&+ \frac{3 F_1 \beta_3}{32 \Omega_1^3} \left( \cos\theta_{1}(2X^2 + A^2) - 2 X Y \sin\theta_{1} \right)
+ \frac{F_{1}}{8 \Omega_1^3} \left( \Gamma \Omega_1 \sin\theta_{1} - 5 \Omega_1^2 \cos\theta_{1} + \Omega_0^2 \cos\theta_{1} \right).
\end{align}

These equations show that the dominant (mechanical-amplitude-dependent) contributions from the quadratic nonlinearity $\beta_{2}$ in the second order can be expressed as a renormalized cubic nonlinearity. Namely, the terms containing $\beta_{2}^2 A^2$ in equations~\ref{eq:duffing_two_drive_slowflow_2nd} can be written as $\beta\rightarrow \beta_{3}-\frac{10}{9}\frac{\beta_{2}^2}{\Omega_{1}^2}$, as in \eqref{eq:duffing_two_drive_slowflow}. The second-order shift from the quadratic term is already included in the experimental $\beta$, consistent with the excellent theory–experiment match. Thus, no additional corrections are required. Finally, note that this analysis holds even in the presence of extra drive tones.

\subsection{Numerical Simulation}
To reproduce the experimental protocol in the simulations, we initialize the system in the low-amplitude solution of the single-tone driven Duffing oscillator, i.e $F_2=0$. We use HarmonicBalance.jl \cite{kosata2022harmonicbalance} to define the equations of motion for the Duffing given by Eqs. \eqref{eq:duffing_two_drive_slowflow} and solve them to obtain the low-amplitude solution for the fitted experimental parameters, setting $F_2=0$. At $t=0$, we `switch-on' the second drive $F_2=hF_1$, where the rate of change of the phase $\varphi(t)$ depends on the probe detuning $\Delta_{21}$, $\frac{\mathrm{d} \varphi(t)}{\mathrm{d} t}=\Delta_{21}$. We use HarmonicBalance.jl's wrapper around DifferentialEquations.jl \cite{rackauckas2017differentialequations} to solve the coupled differential equations given by Eqs. \eqref{eq:duffing_two_drive_slowflow} using an explicit Runge-Kutta integrator. This gives us the quadratures $X(t)$ and $Y(t)$ and the resultant phase space trajectories of the system, as shown in Fig. 4 of the main text.

The steps involved in the implementation using HarmonicBalance.jl are: 
\begin{enumerate}
   \item We choose a final phase $\phi_f=10\pi$ to neglect any transient effects and ensure that the system evolves for complete cycles.
   \item For a particular detuning $\Delta_{21}$, we calculate the final time for the evolution as $t_f=\frac{\phi_f}{|\Delta_{21}|}$. This final time is used to define a linear sweep of the parameter $\varphi (t)$ in $[0, 10\pi]$ over the time-span $[0, t_f]$. This enables us to use the functionalities of HarmonicBalance.jl \cite{kosata2022harmonicbalance} to solve Eqs. \eqref{eq:duffing_two_drive_slowflow} using DifferentialEquation.jl.   
   \item We obtain the solution for $X(t), Y(t)$ which can be used to plot the system's trajectory in phase space, as shown in Fig. 4 of the main text.
 
\end{enumerate}

\end{widetext}
%


\end{document}